\def\USEACHEMSO{0} %

\if\USEACHEMSO1
\documentclass[journal=jctcce,manuscript=article,%
email=false,%
]{achemso}
\else
\documentclass[aip,jcp,notitlepage,
tightenlines,
twocolumn,
reprint,%
floatfix]{revtex4-2} 
\fi

\usepackage[utf8]{inputenc}
\usepackage[T1]{fontenc} 
\usepackage{stix}
\usepackage[version=3]{mhchem}
\usepackage{hyperref}
\usepackage{units}

\usepackage{booktabs}
\usepackage{graphicx}
\usepackage[table]{xcolor}
\usepackage{array}
\usepackage{dcolumn}%
\usepackage{comment}
\definecolor{Gray}{gray}{0.85}

\usepackage{xspace}
\usepackage{verbatim}
\let\oldtheequation\theequation
\makeatletter
\def\tagform@#1{\maketag@@@{\ignorespaces#1\unskip\@@italiccorr}}
\renewcommand{\theequation}{(\oldtheequation)}
\makeatother 

\if\USEACHEMSO1
\fi

\usepackage{listings}
\usepackage[pagewise]{lineno}%
\let\oldequation\equation
\let\oldendequation\endequation

\renewenvironment{equation}
  {\linenomathNonumbers\oldequation}
  {\oldendequation\endlinenomath}
\let\oldalign\align
\let\oldendalign\endalign

\renewenvironment{align}
  {\linenomathNonumbers\oldalign}
  {\oldendalign\endlinenomath}

\usepackage[textsize=small,textwidth=2.2cm]{todonotes}
\newcommand{\angstrom}{\textup{\AA}} %
\newcommand{\matr}[1]{\ensuremath{\mathbf{#1}}}
\newcommand{\braket}[2]{\ensuremath{ \langle #1 | \, #2  \rangle }}

\newcommand{\ketbra}[2]{\ensuremath{  | {#1} \rangle \langle {#2} |}}

\newcommand{\ket}[1]{\ensuremath{  | {#1} \rangle}}

\newcommand{\matrixe}[3]{\ensuremath{ \langle{#1} | {#2} | {#3} \rangle }}

\newcommand{\cre}{\hat a^\dagger}
\newcommand{\ann}{\hat a}

\newcommand{\Nel}{\ensuremath{N_\text{el}}}

\newcommand{\Ncluster}{\ensuremath{C}}
\newcommand{\Norb}{\ensuremath{K}}
\newcommand{\NorbCAS}{\ensuremath{K_\text{act}}}
\newcommand{\NorbExt}{\ensuremath{K_\text{ext}}}
\newcommand{\NorbThawed}{\ensuremath{K_\text{inact}}}
\newcommand{\NorbInt}{\ensuremath{K_\text{Int}}}
\newcommand{\NdetRef}{\ensuremath{n_\text{ref}}}
\newcommand{\Ndet}{\ensuremath{P}}
\newcommand{\NdetExt}{\ensuremath{P_\text{ext}}}
\newcommand{\bdim}{\ensuremath{D}}
\newcommand{\bdimCAS}{\ensuremath{D_\text{act}}}
\newcommand{\bdimExt}{\ensuremath{D_\text{ext}}}

\newcommand{\lit}[1]{Ref.~[\!\!\citenum{#1}]\xspace}
\newcommand{\lits}[1]{Refs.~[\!\!\citenum{#1}]\xspace}

\definecolor{CBred}{RGB}{215,25,28}
\definecolor{CBdblue}{RGB}{5,113,176}
\newcommand{\new}{}

\if\USEACHEMSO0
\begin{document}
\fi

\title{Matrix product states with large sites}
\author{Henrik R.~Larsson}
\email{larsson [a t] caltech . e$\delta$u}
\affiliation{Division of Chemistry and Chemical Engineering, California Institute of Technology, Pasadena, CA 91125, USA}

\author{Huanchen Zhai}
\affiliation{Division of Chemistry and Chemical Engineering, California Institute of Technology, Pasadena, CA 91125, USA}

\author{Klaas Gunst}
\if\USEACHEMSO0
\altaffiliation[Present address: ]{Quantum Simulation Technologies, Inc., Cambridge, MA 02139}
\fi
\if\USEACHEMSO1
\altaffiliation{Present address: Quantum Simulation Technologies, Inc., Cambridge, MA 02139}
\fi
\affiliation{Division of Chemistry and Chemical Engineering, California Institute of Technology, Pasadena, CA 91125, USA}
\affiliation{Center for Molecular Modeling, Ghent University, Technologiepark 46, B-9052 Zwijnaarde,
Belgium}
\affiliation{Department of Physics and Astronomy, Ghent University, Krijgslaan 281, S9, B-9000
Ghent, Belgium}

\author{Garnet Kin-Lic Chan}
\email{garnetc [a t]  caltech . e$\delta$u}
\affiliation{Division of Chemistry and Chemical Engineering, California Institute of Technology, Pasadena, CA 91125, USA}

\if\USEACHEMSO1
\begin{document}
\fi

\begin{abstract}
We explore various ways to group orbitals into clusters in a matrix product state (MPS). 
We explain how a generic cluster MPS can often lead to an increase in computational cost and instead propose a special cluster structure, involving only the first and last orbitals/sites, with a wider scope for computational advantage. 
This structure is a natural formalism to  describe correlated multireference (MR) theories. 
We demonstrate the flexibility and usefulness of this approach by implementing various uncontracted MR configuration interaction, perturbation and linearized coupled cluster theories using an MPS with large cluster sites.
Applications to the nitrogen dimer, the chromium dimer, and benzene, including up to triple excitations in the external space, demonstrate the utility of an MPS with up to two large sites. 
We use our results to analyze the quality of different multireference  approximations.
\end{abstract}

\maketitle
\section{Introduction}
\label{sec:intro}
\begin{figure}[!tbp]
  \includegraphics[width=.9\columnwidth]{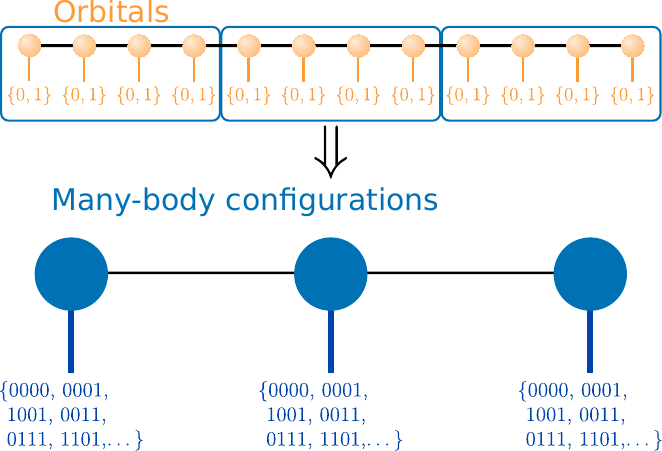}
    \caption{A cluster matrix product state  obtained by combining multiple orbitals into large sites. The Hilbert space of each of the large sites consists of $\Ndet$ many-body configurations that may be further approximated.\cite{asd2014parker}}
  \label{fig:clusterMPS}
\end{figure}

The density matrix renormalization group (DMRG) and its associated ansatz of matrix product states (MPSs) \cite{dmrgRev2011schollwock,dmrg1992white,dmrg1993white,dmrgRev2011chan,dmrgRev2020reiher,dmrgRev2015legeza}
are established as useful electronic structure approximations in problems where there are a large number of correlated open shells.%
\cite{Density2008martia,Highperformance2009kurashige,
Entangled2013kurashige,Lowenergy2014sharma,Electronic2019li,Electronic2019lia,Multireference2019roemelt}
The formalism requires first mapping the orbitals to a one-dimensional lattice of sites. %
One direction discussed already in the first quantum chemistry DMRG paper\cite{qcDMRG1999white}
is the possibility of grouping clusters of related orbitals into large ``sites'' (\autoref{fig:clusterMPS}), whose Hilbert space is then approximated outside of the  truncation procedure of DMRG. Such a cluster MPS seems attractive for interpretation, as one can group orbitals corresponding to chemical identity, 
and it has been \new{efficiently} implemented in a number of works.\cite{asd2014parker,comb2021li,ttns2013nakatani,larsson2020mmps,tpsci2020abraham,Rankone2019nishio}
However, because the Hilbert space of the cluster site grows exponentially quickly with the number of cluster orbitals, the computational advantage is less clear. \new{In addition, the drawbacks of clustering, which gives rise to more complicated interactions and entanglement between clusters, has been understood since the earliest formulations of the quantum renormalization group.\cite{dasgupta1981real,white1992real}
In the first part of this work, we analyze whether clustering orbitals is a good idea in chemical problems from the view of computational cost and accuracy.}

\begin{figure}%
  \includegraphics[width=.9\columnwidth]{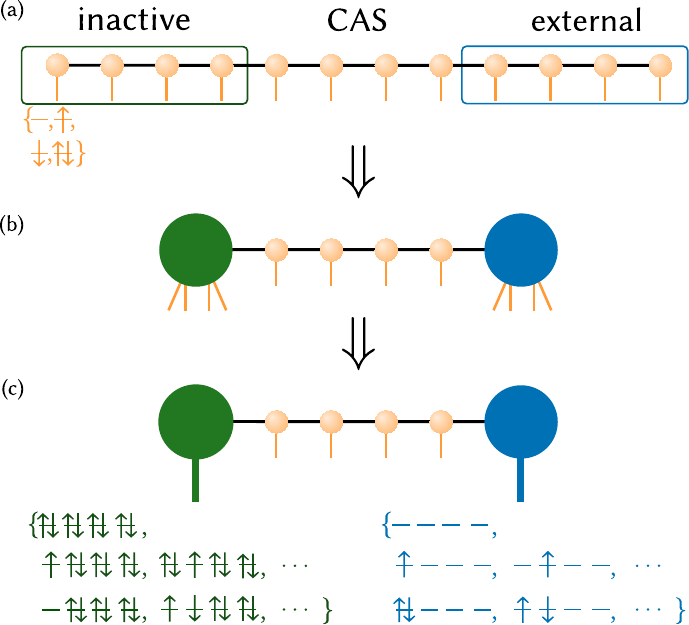}
    \caption{Panel (a), (b): Diagrammatic representation of a cluster MPS with two large sites at either end of the lattice. This ansatz demonstrates favourable scaling with respect  to the large site Hilbert space dimension because there is no shared boundary between the large sites. 
    Panel (c): Application to 
    an uncontracted multireference dynamic correlation wavefunction in the singles and doubles space, where the Hilbert space of the green site covers the inactive
    orbitals and the Hilbert space of the blue site covers the external orbitals.
    }
  \label{fig:largeSite}
\end{figure}

In the second part, we discuss a specific setting where grouping sites into large clusters has a clear theoretical computational advantage.  This occurs when the clusters are at either end of the DMRG lattice (\autoref{fig:largeSite}). 
Because the clusters do not share a common boundary, 
the cluster Hilbert space dimension appears together with the MPS bond dimension in a computationally more favourable way than in a general cluster MPS. 
A natural application for this type of MPS is to represent dynamical correlations by clustering inactive and external orbitals. 
As we demonstrate, this leads to a substantial cost reduction in MPS treatments of dynamic correlation.

The paper proceeds as follows. We start by analysing the cost of cluster MPS (\autoref{sec:theory_clustering}) and
explain why computational gains are not expected in general settings. 
We then analyse the conditions leading to the favourable cost of the single and double cluster-site model (\autoref{sec:theory_single_large_site}) and discuss its application to uncontracted multireference dynamic correlation theories (\autoref{sec:theory_fois_mps}).
We next discuss the detailed implementation of DMRG with large cluster sites in \autoref{sec:theory_impl}, taking advantage of the large-scale parallel DMRG implementation in \lit{zhai2021Low}. 
Finally in \autoref{sec:applications}, we demonstrate the cluster MPS implementation of uncontracted multi-reference configuration interaction (MRCI), 
multi-reference perturbation theory (MRPT), and multi-reference linearized coupled cluster theories (MRLCC) in applications to the nitrogen dimer, chromium dimer, and the benzene molecule, using complete active spaces (CASs) with up to 30 electrons and 30 orbitals, with up to triples in the external space, and with up to 280 external orbitals. 
We conclude in \autoref{sec:conclusions}.

\new{Note: while this manuscript was under review, a related preprint appeared that implements uncontracted MRCI by using an MPS with a single cluster site.\cite{Largescale2021barcza}} 
\section{Theory}
\label{sec:theory}
\subsection{Analysis of clustering sites in matrix product states}
\label{sec:theory_clustering}
The benefit of clustering orbitals depends on the entanglement structure of the problem. 
If the entanglement is such that groups of sites are strongly entangled internally, 
but only weakly entangled between the groups, 
 then it may make sense computationally to cluster into large sites. The critical question is how large this difference in intra- versus inter-cluster entanglement needs to be for a computational benefit.

To start, we recall the computational cost of the standard DMRG algorithm, then examine the cost for multiple clusters of sites, and then finally for large sites at the ends of the DMRG lattice, the latter being the main focus in this work.

\subsubsection{Standard matrix product states}
\label{sec:theory_dmrg}
In the standard MPS/DMRG formulation, the wavefunction for $\Norb$ orbitals is
mapped to a lattice of $\Norb$ sites and written as a
 matrix product state (MPS) of bond dimension $\bdim$,\cite{dmrgRev2011schollwock}
\begin{align}
    |\Psi\rangle &= \sum_{ \{ n_i \}} \matr{A}^{n_1} \matr{A}^{n_2} \ldots \mathbf{A}^{n_\Norb} \ket{n_1 n_2 \ldots n_\Norb},
\end{align}
where the matrices $\matr{A}^{n_i}$ at ``site'' $i$ 
are of dimensions $\bdim_{i}\times \bdim_{i+1}$, 
save those associated with the first and last sites,
which are vectors of size $\bdim_i$. 
The bond dimension of the MPS is then defined as  $\bdim=\text{max}_i \bdim_i $. 
The local Hilbert space $\{ \ket{n} \}$ is that of a spatial orbital $\phi$ 
and the dimension is $\Ndet=4$
($\{ \ket{n_i} \} = \{\ket{\mathrm{vac}}, \ket{\phi_i^\alpha}, \ket{\phi_i^\beta}, \ket{\phi_i^\alpha \phi_i^\beta}\}$). 

The main cost of the DMRG algorithm when using electronic structure Hamiltonians stems from two steps, performed at each site:
(1) the construction and diagonalization of an effective Hamiltonian in 
the product space of one or more sites and the
renormalized Hilbert space of their environment, (2) the transformation of operators into the new renormalized space.\cite{qcDMRG1999white,dmrg2002chan}
Both steps contribute to the leading computational scaling, which is usually given (per site) as
$\mathcal O[(\Norb_<^2+\Norb) \bdim^3\Ndet^2 + \Norb_<^2 \Norb_>  \bdim^2]$, assuming $D_i \sim D$. $\Norb_<$ ($\Norb_>$) is the smaller (larger) of the numbers of orbitals to the left and right of the bipartition at site $i$.
In the following we assume that $\bdim \gg \Norb$ in order to drop the second term, which stems from the transformation of operators.
In addition, as $\Ndet=4$ is a small constant, we drop the $\Ndet$ dependence for a standard MPS.
For the total cost, $\Norb_{</>}\sim \Norb$ and  
the cost per site is then multiplied by $\Norb$  to obtain the leading cost for the DMRG algorithm of $\mathcal O[\Norb^3 \bdim^3]$. 
For reference, the precise scalings for a normal MPS and different variations of MPSs discussed below are  gathered in  \autoref{tab:scaling}.

\begin{table*}[!htbp]
\caption{Comparison of the scaling for the computational cost of the various MPS-based DMRG formulations. 
$\bdim$ defines the bond dimension, $\Norb$ the number of orbitals, $\Ndet$ the number of configurations on the large site, and $\Ncluster$ the number of clusters (large sites) in a cluster MPS.  For the MPS with large sites at either ends, we here assume only one large site with $\NorbExt$ orbitals, for simplicity. The number of standard (orbital-based) sites is then $\NorbCAS$ and $\NorbExt + \NorbCAS=\Norb$. The first column shows the costs for carrying out the Davidson diagonalization and operator renormalization steps, while the second column is an additional cost for the complementary operator renormalization step.
}
\label{tab:scaling}
\begin{tabular}{lcc}
\toprule
  method & diagonalization \& renorm.~operators &  compl.~renorm.~op.\\
  \midrule %
  \rowcolor{Gray}
  normal MPS&%
  $\Norb^3\bdim^3$ &%
  $\Norb^4 \bdim^2$  \\
cluster MPS& %
$\Ncluster \Norb^2 [\bdim^3 \Ndet +  %
\bdim^2 \text{poly}(\Ndet)]$ &%
$\Ncluster \Norb^3  \bdim^2$\\ %
  \rowcolor{Gray}
single large site MPS& $\NorbCAS (\NorbCAS^2 + \Norb) \bdimCAS^3$ & $\NorbCAS^2 \Norb \bdim^2$\\
  \rowcolor{Gray}
& $+ (\NorbCAS^2 + \Norb) [\bdimExt^2 \Ndet + \bdimExt\text{poly}(\Ndet)] $ &
\\
MPS-MRCISD (large site MPS)&  as single large site MPS with $\Ndet = \NorbExt^2$ &\\
  \rowcolor{Gray}
MPS-MRCISD & $\NorbCAS (\NorbCAS^2 + \Norb) \bdimCAS^3$ & $\NorbCAS^2 \Norb \bdim^2$\\
  \rowcolor{Gray}
(conventional MPS)& $+ \sum_{i=\NorbCAS}^{\Norb} (\Norb_{<,i}^2 + \Norb) \bdim_i^3$ & $+ \sum_{i=\NorbCAS}^{\Norb} \Norb_{<,i}^2\Norb_{>,i} \bdim_i^2$\\
  \bottomrule
\end{tabular}
\end{table*}
\subsubsection{Cluster matrix product states}
\label{sec:theory_cluster_mps}

The above analysis can be repeated for a cluster MPS with clusters of orbitals as sites. %
We assume that each cluster has $\Norb_c$ orbitals, with a cluster Hilbert space of $\Ndet$ configurations. 
Then the number of sites in the cluster MPS is reduced to $\Norb/\Norb_c = \Ncluster$. 
To simplify the analysis, we assume that $\Norb_c$ is similar for all clusters.
Note that to obtain the same accuracy as the standard MPS %
the bond dimension used  between two clusters must be the same as the bond dimension in the standard MPS between the sites at the boundary of the two clusters. 
Now consider increasing the number of orbitals in each cluster $\Norb_c$.  %
Assuming the full Hilbert space of each cluster is used, 
then $\Ndet \sim 4^{\Norb_c}$. Because $P$ is now potentially large, we consider it as important as $D$ and $K$ in the analysis of the leading scaling.

The cost of the cluster MPS is then given by  $\mathcal O\{
\Ncluster \Norb^2 [\bdim^3 \Ndet + \bdim^2 \text{poly}(\Ndet)] + \Ncluster \Norb^3  \bdim^2\}$, %
analogous to the standard MPS cost.
We again assume that $\bdim \gg \Norb$ and drop the last term.
In contrast to the cost given above for the standard MPS, here we have written the first term without a $\Ndet^2$ dependence because we assume the use of a tri-partition to perform the diagonalization and renormalization. 
As explained below in \autoref{sec:impl_cluster_mps},
this changes the cost of the first term per site from $\mathcal O(\Norb_<^2 D^3 \Ndet^2)$ to $\mathcal O(\Norb_< \Norb_> D^3 \Ndet)$. %
The term containing $\text{poly}(\Ndet)$ stems from applying operators in the cluster space onto the site and is at most $P^2$. However, 
in many common situations we can use a local basis (such as a determinantal basis) in which the Hamiltonian is sparse.
Then $\text{poly}(\Ndet)\sim \Ndet$. Hence, the leading cost of the cluster MPS simplifies to  $\mathcal O(\Ncluster \Norb^2 \bdim^3 \Ndet)$.

While $\Ndet$ only appears linearly in the scaling, it grows much faster than the $1/\Norb_c$ reduction in the number of sites in the cluster MPS. 
Although there are ways to truncate the cluster Hilbert space, e.g., via filtering determinants (``selected configuration interaction''),\cite{asd2014parker} general linear subspace projection (``Tucker decomposition''),\cite{Tensor2009kolda} 
or using an additional factorized ansatz for the MPS cluster matrix (``comb tensor networks''),\cite{comb2019chepiga,comb2021li} any choice must achieve an effective exponential reduction in $P$ complexity, to compete with the standard MPS cost. This means that in systems whose entanglement is well described by a standard MPS with constant bond dimension across the lattice, a cluster MPS is unlikely to reduce computational cost for the same accuracy.

A different limiting case is in problems described by an MPS with highly non-uniform bond dimensions, large within a cluster of sites and very small between clusters.
The extreme case is no entanglement ($\bdim=1$) between clusters, i.e.~the state is a product state of cluster wavefunctions (such as a generalized valence bond wavefunction, or for infinitely separated or noninteracting systems).\cite{larsson2020mmps,Rankone2019nishio,cMF2015jimenez-hoyos,Multiconfigurational2019hermes}
Since $\bdim=1$, it does not appear in the scaling and we need to consider terms non-leading in $\bdim$ in the analysis. Assuming determinant-like sparsity in the local basis,
the DMRG cost is then $\mathcal O[\Ncluster (\Norb^2 \Ndet+\Norb^3)]$.  %
Conversely, when treating the problem using a standard MPS, $D > 1$ when cutting across a cluster. In the worst case, $\bdim \sim \mathcal O(\Ndet^{1/2})$, yielding $\mathcal O(C K^2 \Ndet^{3/2})$ cost (the leading $D$ term).
This is larger than the cluster MPS result because we assumed no compressibility within the cluster, and the renormalized MPS basis cannot use sparsity. Thus the cluster MPS is advantageous in this limit.

In general, chemical problems fall between these two limiting cases. 
Sufficiently weakly interacting units are close to the second limiting case, and thus there are computational benefits to using the cluster MPS there. %
 But the exponential overhead of clustering, together with the presence of long-range interactions (which introduce long-range entanglement) means that many problems are in fact close to the first scenario.
To illustrate this, in Appendix \ref{sec:appl_H8} we carry out numerical simulations using both cluster MPS as well as the ordinary MPS for hydrogen chains at several geometries. The results show
that there can be little gain from clustering even in a regime where the chemical identity of individual atoms or molecular units is evident.

\subsection{Matrix product states with large sites at the ends}
\label{sec:theory_single_large_site}
We now turn to the case of main interest in this work,
when there are large sites at one or both ends of the MPS lattice. 
In anticipation of the multireference use-cases discussed later, the orbitals treated in the usual MPS fashion will be denoted active orbitals, orbitals in the left cluster will be denoted inactive, orbitals in the right cluster will be denoted external. The number of orbitals in each class is then $\NorbCAS$, $\NorbThawed$, $\NorbExt$ respectively.
The ansatz thus takes the form (see also \autoref{fig:largeSite})
\begin{equation}
\begin{split}
    |\Psi\rangle =& \sum_{ \{ n \}} \matr{A}^{\matr{n}_\text{inact}}  \mathbf{A}^{n_{\NorbThawed+1}} \mathbf{A}^{n_{\NorbThawed+2}} \ldots %
    \mathbf{A}^{\matr{n}_{\text{ext}}}\\%
   &\qquad \ket{\matr n_\text{inact} n_{\NorbThawed+1} n_{\NorbThawed+2} \dots %
    \matr n_\text{ext}},\label{eq:singleLargeSiteMPS}
\end{split}
\end{equation}
where $\matr{n}_\text{inact}$ and $\matr{n}_\text{ext}$ label the Hilbert space of the inactive and external sites, of dimension $\Ndet_\text{inact}$, $\Ndet_\text{ext}$ respectively.

To simplify the scaling discussion, we will ignore the left large site, i.e., $\NorbThawed=0$. The main finding is easily generalized to the case of  $\NorbThawed\neq0$. 
Following the discussion above, the DMRG cost at sites within the active space is the same as in standard DMRG, i.e.~$\mathcal O[(\NorbCAS^2+\Norb) \bdim^3 \Ndet^2]$ ($P=4$), where the only difference is that $\NorbCAS \neq\Norb$.
The new consideration is for the site at the boundary between the active sites and the large external site.
The contraction at the boundary has cost $\mathcal O[(\Norb_\text{act}^2 + \Norb) [\bdimExt^2 \Ndet_\text{ext} +\bdimExt\text{poly}(\Ndet_\text{ext})]$ 
where $\bdimExt$ is the bond dimension at the boundary. 
As for the cluster MPS, we assume  $\text{poly}(\Ndet_\text{ext}) \sim \Ndet_\text{ext}$ and drop the last term.
Unlike in the general cluster MPS, however, the cluster Hilbert space dimension appears with $\bdimExt^2$, not $\bdimExt^3$. Thus, for $P_\text{ext}$ not too large (see below) it is possible to obtain a speedup. For the case of two large sites we must also consider the boundary between the inactive large site and the active space, but this takes the same form where the inactive cluster Hilbert space dimension is multiplied by $D_\text{inact}^2$.
In the limiting case of $\NorbCAS=0$, i.e.~the MPS consists of only two large sites, the cost scales as $\mathcal O[ (\NorbInt^2 + \NorbExt) \bdimExt^2 \Ndet_\text{ext}]$. This corner case may be advantageous when $\Ndet_\text{ext}$ is small enough, but will not be considered further here.

One concrete application is to use the ansatz \autoref{eq:singleLargeSiteMPS} %
to represent orbital partitioned quantum chemistry models, such as the restricted active space (RAS) model and other uncontracted multi-reference dynamic correlation models. 
For example, if we assume a singles and doubles theory where the external space contains at most two electrons, then $\NdetExt \sim K_\text{ext}^2$. 
Using a standard MPS to represent such a state, the external space restriction limits the bond-dimension of the MPS to $\bdimExt$ at the boundary. The cost of the standard DMRG contraction at the boundary site is $\mathcal{O}(\NorbExt^2  \bdimExt^3\Ndet^2)$ (with $\Ndet=4$), and assuming $\bdimExt$ decreases linearly across the external orbitals, the leading cost becomes $\mathcal O(K_\text{ext}^3 \bdimExt^3)$ (the precise scaling is detailed in \autoref{tab:scaling}).
However, using a large external site, and the expression for the single boundary contraction, then for $\Norb_\text{ext} \sim \Norb > \Norb_\text{act}^2$ we obtain a speedup of $\bdimExt$ relative to the standard DMRG implementation. 
If $\NorbExt < \NorbCAS^2$, the speedup will be larger than $\sim \bdimExt / \NorbExt$.
For an external space with only single excitations out of the active space,
the speedup is even greater, namely up to $\Norb \bdimExt$.
For a more general external space, e.g., constructed by selected configuration interaction,\cite{Iterative1973hurona,asci2016tubman,sci2016schriber,HeatBath2016holmes} a similar analysis can be applied.

\subsection{Matrix product state formulation of uncontracted dynamical correlation methods}
\label{sec:theory_fois_mps}

We next describe how to approximate various uncontracted multireference dynamical correlation methods using MPS. In all these cases, the large site MPS ansatz~\ref{eq:singleLargeSiteMPS}  
can be used, and when the excitation degree is small (e.g., up to singles and doubles, in some cases up to triples) we can expect speedup relative to the standard MPS formulation. 
This will be assessed in the benchmark in \autoref{sec:applications}.%

\subsubsection{Multireference configuration interaction theory}
The uncontracted multi-reference CI (uc-MRCI) ansatz takes the form
\begin{align}
\ket{\Psi_\text{uc-MRCI}} &= 
\label{eq:mrcisd_wf}
 \sum_{C}^{\NdetRef} c_C \ket{C} + \sum_{E} c_E \ket{E} 
\end{align}
where $\ket{C}$ denotes a configuration from the reference space (no particles in the external space, no holes in the inactive space) and $\ket{E}$ are configurations outside of the reference space, classified as singles (one particle in the external space), doubles (two particles in the external space) and so on. The uc-MRCI coefficients $c_C$ and $c_E$ are determined by minimizing the variational energy. 

MRCI does not give an extensive energy, e.g., the energy of independent subsystems is not the sum of the energy of the systems. Defining the correlation energy as  $\Delta E = E_\text{MRCI} - E_0$ ($E_0$ being the energy of the reference wavefunction $\ket{\Psi_0}=\sum_C c_C^0|C\rangle$, with $c_C^0$ determined variationally), the following approximate size-extensivity corrections have been defined (among others\cite{szalay2012rev}):
\begin{align}
E_\text{D} &= (1-c_0^2) \Delta E,\\
E_\text{RD} &= (1-c_0^2)/c_0^2 \Delta E,\\
E_\text{P} &= \frac{\sqrt{ \Nel^2 + 2 \Nel \tan(2\theta)} - \Nel}{ 2 [\mathrm{sec}(2\theta) - 1]} \Delta E \\
&\approx (1-2/\Nel) E_\text{RD},\qquad \theta = \arccos(c_0),\nonumber\\
E_\text{M} &= g_\text{M}  E_\text{RD},\qquad g_\text{M} = \frac{(\Nel-2) (\Nel-3)}{\Nel (\Nel-1)},
\end{align}
where $E_\text{D}$ ($E_\text{RD}$) is the (renormalized) Davidson correction,\cite{langhoff1974dav,szalay2012rev} 
$E_\text{P}$ is the Pople correction,\cite{pople1977corr}
and
$E_\text{M}$ is the Meissner correction.\cite{meissner1988mrci}
$\Nel$ is the number of correlated electrons and 
$c_0^2$ is either defined as\cite{butscher1977mrciDav}
\begin{equation}
  c_0^2 = \sum_{C}^{\NdetRef} c_C^2,\label{eq:c0_rot}
\end{equation}
or as\cite{blomberg1983mrci} 
\begin{equation}
  c_0^2 = \braket{\Psi_0}{\Psi_\text{MRCISD}}^2.\label{eq:c0_fix}
\end{equation}
Here, we use \autoref{eq:c0_rot}, which has been found to be slightly more accurate in many situations.\cite{szalay2012rev,khait2010cisdtq} 
The size-consistency-corrected MRCI methods will be referred to as MRCI+Q$_X$, where $X$ stands for the particular correction used. 

One can also define an energy functional to variationally minimize that includes the size-extensivity correction in its definition. This permits a simple implementation of the gradients and properties. 
Many such functionals can be obtained by shifting the diagonal of the MRCI Hamiltonian according to
\begin{align}
  \hat H &\to  \hat H + \Delta \hat P,\quad \Delta = (1-g) \Delta E,\label{eq:aqcc_eq}%
\end{align}
where 
\begin{equation}
 \hat P=\sum_E|E\rangle \langle E|, \label{eq:ex_proj}
\end{equation}
and the
parameter $g$ defines the type of correction.\cite{szalay2012rev} Here, we will use only
two variants $g=1-g_M$ (MR averaged quadratic coupled-cluster, MR-AQCC, method,\cite{szalay1993AQCC})
and $g=2/\Nel$ (MR averaged coupled pair functional, MR-ACPF).\cite{gdanitz1988ACPF}
MR-AQCC is related to $E_M$ and MR-ACPF is related to $E_P$, and MR-ACPF is extensive for identical subsystems. 

The MPS versions of the above theories are easily defined, by constraining MPS to preserve  constraints in the inactive and external Hilbert spaces. We will refer to the MPS versions of the above theories by prepending MPS to the name of the method, e.g., MPS-MRCI, MPS-ACPF, MPS-AQCC, etc.
For brevity, we avoid the additional ``uc'' prefix and assume that ``MPS-MR$X$'' implies an uncontracted multireference formulation of method $X$.
\subsubsection{Multireference perturbation theory and multireference coupled cluster theory}
It is straightforward to formulate uncontracted multi-reference perturbation theory in terms of MPS.
This was discussed in \lit{sharma2014Hylleraas} with subsequent extensions in \lits{sharma2015REPT,Quasidegenerate2016sharma,sharma2017icPT,guo2018pDMRG}.
Given a zeroth-order Hamiltonian, $\hat H_0$, the first order perturbed wavefunction, $\ket{\Psi_1}$, can be obtained by minimizing the Hylleraas functional\cite{helgaker_book} 
\begin{equation}
\begin{split}
H[\ket{\Psi_1}] = &  
\matrixe{\Psi_1}{\hat H_0 - E_0}{\Psi_1} \\
&+ 2 \matrixe{\Psi_1}{\hat Q (\hat H-\hat H_0)}{\Psi_0},
\end{split}
\end{equation}
where $\hat Q = 1 - \ketbra{\Psi_0}{\Psi_0}$.
The MPS formulation corresponds to representing both $\ket{\Psi_0}$  and $\ket{\Psi_1}$ as MPS, approximating
the uncontracted perturbation solution. 
The energy for third-order MRPT can be obtained from
\begin{equation}
E_\text{PT3} = E_\text{PT2} + \matrixe{\Psi_1}{\hat{H} - \hat{H}_0}{\Psi_1}
\end{equation}

In \lit{sharma2014Hylleraas} the above theory was implemented for the Dyall Hamiltonian\cite{dyall_1995} 
to approximate the uncontracted $n$-electron valence-state perturbation theory (NEVPT2),\cite{nevpt2001angeli,nevpt2001angeliLett,angeli_2007} and
the resulting formulation was termed MPS-PT2.
The Dyall Hamiltonian $\hat H_{0,D}$ is defined as\cite{dyall_1995,angeli_2007}
\begin{align}
\hat H_{0,D} =& \sum_{ij\in \text{inactive}} F_{ij}\cre_i \ann_j + %
\sum_{rs \in \text{ext}} F_{rs} \cre_r \ann_s + %
\sum_{ab\in \text{act}} h_{ab}^\text{eff} \cre_a \ann_b \nonumber \\ &+ 
\frac12 %
\sum_{abcd\in \text{act}} \braket{ab}{cd} \cre_a\cre_b  \ann_d \ann_c + E_D,\\
h_{ab}^\text{eff} =& h_{ab} + \sum_{i\in\text{inactive}} [2\braket{ai}{bi} - \braket{ai}{ib}],\\
E_D =& 2 \sum_{i\in\text{inactive}} (h_{ii} - F_{ii})  + \sum_{ij\in\text{inactive}} [2\braket{ij}{ij} - \braket{ij}{ji}],
\end{align}
where $\matr F$ is the generalized Fock matrix. 
$h_{ij}=\matrixe{i}{h}{j}$ are the one-particle Hamiltonian matrix elements and $ \braket{ab}{cd}$ the electron repulsion matrix elements.

In \lit{sharma2015REPT}, Fink's restraining the excitation degree (RE) Hamiltonian,\cite{fink2006rept,fink2009MRrept}
\begin{equation}
\begin{split}
\hat H_{0,F} =& E_0 + \sum_{pq; \Delta n_\text{ex} =0} h_{pq}\cre_p\ann_q \\ &%
+ \frac12 \sum_{pqrs; \Delta n_\text{ex}=0} %
\braket{pq}{rs} \cre_p\cre_q\ann_s\ann_r,
\end{split}
\end{equation}
was used, leading to an MPS-based version of RE perturbation theory (REPT).
$\Delta n_\text{ex}=0$ indicates that excitations between the inactive, active, and external spaces are omitted, compared to the full Hilbert space.
If there is no active space, then $\ket{\Psi_0}$ is a single Slater determinant, and the result from REPT2 is identical to the linearized-coupled cluster (LCC) approximation, thus this approximation was termed MPS-LCC in \lit{sharma2015REPT}.
However, given a multi-reference $\ket{\Psi_0}$, this equivalence no longer holds. We will consider
  another linearized multi-reference coupled cluster approximation below, thus we will refer to this choice of $\hat H_{0,F}$ as MPS-MRREPT2. 

\lits{laidig1984lcc,laidig1987lcc} defined the first linearized multi-reference coupled cluster approximation (MR-LCCM). This corresponds to the choice
\begin{align}
H_{0, LCCM} = \hat{P} \hat{H} \hat{P} + \ketbra{\Psi_0}{\Psi_0} E_0  \ketbra{\Psi_0}{\Psi_0}\label{eq:h0_lccm}
\end{align}
where $\hat P$ is defined in \autoref{eq:ex_proj} and $|\Psi_1\rangle$ is solved for only in the excited space, i.e.~$|\Psi_1\rangle = \hat{P}|\Psi_1\rangle$. It differs from REPT2 in 
that (a)  $\ket{\Psi_1}$ %
has no contributions in the reference space
and 
(b)  the excitation spaces of degree $n_\text{ex}>0$ are coupled in $\hat{H}_0$.
We refer to the MPS implementation of this theory as MPS-MRLCCM.

\section{Implementation}
\label{sec:theory_impl}

We have implemented the modified MPS algorithms described above in several ways. We have implemented uncontracted dynamical correlation methods within a standard MPS formulation by restricting the
occupancy of different spaces, as described in \autoref{sec:restrict} within \textsc{Block2}.~\cite{zhai2021Low}
The large site implementation of the dynamical correlation methods is implemented in \textsc{Block2} as well, as described in \autoref{sec:impl_single_large}. 
Finally, the general cluster MPS (used in the computations in Appendix \ref{sec:appl_H8}) is implemented within the DMRG program \textsc{Schwarzbrot}~\cite{larsson2020mmps} as described in~\autoref{sec:impl_cluster_mps}.
For general references on DMRG implementation, we refer to the literature.\cite{dmrgRev2011chan,dmrg2002chan,dmrg2004chan,kurashige2014dmrgRev,olivares-amaya2015dmrgRev,wouters2014dmrgRev,zhai2021Low} The above MPS algorithms are interfaced with the \textsc{PySCF} program.\cite{PYSCF1,PYSCF2} 

\subsection{Restricting configurations in matrix product states} 

\label{sec:restrict}

To implement the unrestricted multireference dynamical correlation theories in \autoref{sec:theory_fois_mps} in a standard MPS, we enforce constraints on the MPS matrices. Elementary symmetries such as particle number or spin symmetry are usually taken into account by introducing irreducible blocks in the MPS site matrices $\matr A^{n_i}$,\cite{dmrgRev2011schollwock,dmrgRev2015legeza,dmrg2002chan,gunst2021Seniority} where each block corresponds to a different symmetry (e.g., number of particles) to the left
and right of the given site.
If the MPS sites are ordered according to the orbital spaces, inactive ($\NorbThawed$), active ($\NorbCAS$), and external ($\NorbExt$), %
the same technique can be used to constrain the MPS ansatz to a wavefunction of the form~\ref{eq:mrcisd_wf} with a given excitation level.
For example, restricting the particle numbers on 
site $i \leq \NorbThawed$ to be $\{i,i-1,i-2\}$ and 
the particle numbers on site $i > \NorbInt$ to be $\{\Nel, \Nel-1, \Nel-2\}$, we approximate the ansatz~\ref{eq:mrcisd_wf} with singles and doubles excitations.%
\footnote{We assume here that the inactive orbitals are placed on the left of the MPS and that the external orbitals are placed on the right of the MPS. The particle number then increases from site to site until the total electron number, $\Nel$ at the end of the last site.
Corner cases are neglected. %
}
Particle number restriction is sufficient to implement the uncontracted multi-reference dynamical correlation approaches in this work, but extensions to  
 other symmetry sectors (e.g., $S_z$ and $S^2$ symmetry) is possible, and can, for example
 be used to describe wavefunctions restricted by the seniority quantum number.\cite{gunst2021Seniority}
In passing, we note that this approach is very different from the ``multilevel'' DMRG,\cite{mldmrg2015ma} where different maximal bond dimensions are used in the three subspaces, without any restrictions on the particle number blocks.

\subsection{Matrix product states with large sites in the ends}
\label{sec:impl_single_large}

Introducing large sites in an MPS requires significant changes in the implementation of a DMRG code. Since for conventional MPSs in electronic structure theory, the physical dimension of a site is $\Ndet =4$, or, with spin-orbital sites, even $2$, one typically does not optimize the DMRG implementation around the size of the physical dimension. 
However, for large sites, $\Ndet$ can become  arbitrarily large. Hence care must to be taken to avoid unfavorable costs and scaling with respect to $\Ndet$.

In standard CI methods, the matrix representation of operators is seldom explicitly constructed, and instead matrix vector products, such as $\hat{H} \ket{\Psi}$ are evaluated on the fly. 
Here, however, we store all required operators that act in the large site Hilbert space and represent them as sparse matrices (in the determinantal basis of the large site) of size $\Ndet \times \Ndet$. 
This is because (a) operators in DMRG need to be accessed more often than in standard CI methods 
and (b) the size of the large site basis is small ($\sim 10^6$), compared to standard CI methods.
Note that the determinantal configurations range over different numbers of electrons (e.g., between $0$-$2$ for the external site in an implementation of the multireference singles and doubles theories) and this yields
extremely sparse operator matrices depending on the operator, e.g., with $\mathcal O(\Ndet)$ or even $\mathcal O(1)$ nonzeros. 
In our implementation, for up to two electrons in the external space, most of the memory and runtime (including the initialization)  is spent on optimizing the regular sites in the active space and not the large inactive or external sites.

A standard DMRG implementation uses a decomposed form of the Hamiltonian $\hat{H} = \sum_{\alpha} \hat{O}_L^\alpha \hat{O}_R^\alpha$ to carry out the optimization of the MPS matrix at a given site, where $\hat{O}_L^\alpha$, $\hat{O}_R^\alpha$ define operators that act to the left/right(inclusive) of the given site. There are multiple such decompositions,\cite{zhai2021Low} e.g., we can group the Hamiltonian integrals with either the left or right operator, resulting in a normal (no integrals) or complementary (with integrals) operator.
When approaching the site in the middle of an MPS during a sweep, 
it is advantageous in the standard algorithm to swap the assignment of normal and complementary  between the left and right operators in order to reduce the number of terms in the $\alpha$ sum.\cite{zhai2021Low,mpo2016chan,dmrg2009kurashige}
In the large site implementation of multireference dynamic correlation, this is not required, because
we can usually assume that $\NorbExt \gg \NorbThawed+\NorbCAS$, and thus the sweep is always over the first "half" of the sites. In practice, this means that only the normal operators are constructed
for the inactive and active sites, and only the complementary operators are constructed for the last external site.

The standard DMRG algorithm extracts a renormalized basis at each site by constructing and diagonalizing a density matrix.\cite{dmrg1992white,dmrg2002chan}
Small perturbations are also added to this density matrix to 
improve convergence during the optimization.\cite{white2005noise}
For the large site, however, the density matrix would have size $\mathcal O(\Ndet^2)$. To avoid constructing this large object,
we %
use the singular value decomposition (SVD) of the large site\new{, $\matr A^{\matr n_\text{ext/inact}}$, } which reduces the scaling of the memory to $\mathcal O(\bdim\Ndet)$.\cite{hubig2015noise,larsson2019ttns}
Finally, standard DMRG simulations often use a two-site algorithm where two adjacent sites are optimized simultaneously, in order to improve convergence and to optimize the distribution of symmetry blocks in each MPS matrix.
We use a two-site algorithm on all sites except  the large sites, which are treated using the one-site algorithm.
To ensure that symmetry sectors in each matrix are not lost during the DMRG sweep, we always retain at least one state in each symmetry sector (particle number, point group, and $S_z$) in the SVD. Our large site implementation does not currently use $S^2$ symmetry.

To evaluate the scalar size-extensivity energy corrections, 
the weight of the reference space, $c_0^2$, is required. An evaluation as \autoref{eq:c0_fix} via $|\braket{\Psi_0}{\Psi_\text{MRCISD}}|^2$ is done by straightforward contraction of two MPSs.\cite{dmrgRev2011schollwock} 
An evaluation as \autoref{eq:c0_rot} via $\sum_C c_C^2$ requires first setting the inactive and external configurations in the MPS ansatz~\ref{eq:singleLargeSiteMPS},
$\matr n_\text{inact}=\{2_1\dots 2_{\NorbThawed}\}$, $\matr n_\text{ext} = \{ 0_1 \dots 0_{\NorbExt}\} $, and then computing the norm by contracting the resulting MPS with itself.
 The energy functionals associated with the AQCC and ACPF methods are implemented by modifying the diagonal of the Hamiltonian during the optimization, as shown in \autoref{eq:aqcc_eq}. 
We shift the  diagonal of $\hat H$, excluding the reference space, by constructing a matrix product operator (MPO)\cite{dmrgRev2011schollwock,mpo2016chan} representation of
$\hat P$ defined in \autoref{eq:ex_proj}, giving $\hat H \to \hat H + \Delta \hat P$.
For $\Delta E$, we evaluate the correlation energy using the lowest energy so far observed during the DMRG sweep.

The perturbation-based methods MPS-MRLCCM and MPS-MRREPT2 are implemented in the DMRG sweep algorithm by solving linear equations at each site instead of eigenvalue problems.\cite{sharma2014Hylleraas}
For MPS-MRLCCM, \autoref{eq:h0_lccm} can be constructed %
by removing the reference space in  $\ket{\Psi_1}$. %
The zeroth-order Hamiltonian in MPS-MRREPT can be constructed by including only particle-number-conserving operators on the large sites. %
\subsection{Cluster matrix product states}
\label{sec:impl_cluster_mps}

While the implementation of the cluster MPS follows that of an MPS with many large sites, more care has to be taken to avoid a $\Ndet^2$-type of computational cost.
In addition to the modifications described in \autoref{sec:impl_single_large}, 
all DMRG optimization sweeps are performed in one-site mode and explicitly blocked operators (which act on the Hilbert space of a block enlarged by a site) are not explicitly constructed. (For example, this means that we always use a tripartition of the Hamiltonian  $\hat{H} = \sum_\alpha \hat{O}^\alpha_L \hat{O}^\alpha_S \hat{O}^\alpha_R$ where the $S$ index denotes the site being optimized in the sweep). %
This changes the $\mathcal O(\Norb_<^2 D^3 \Ndet^2)$ term in the  scaling of operator multiplication at a site in the DMRG sweep to $\mathcal O(\Norb_< \Norb_> D^3 \Ndet)$, c.f.~\autoref{sec:theory_cluster_mps}. 

\section{Applications}
\label{sec:applications}
\subsection{Nitrogen dimer}
\label{sec:appl_N2}
Here, we compare relative timings of (a) a standard DMRG computation (approximating full configuration interaction, FCI),
(b) an MPS-MRCISD computation based on a standard MPS with restricted quantum numbers and %
(c) an MPS-MRCISD computation based on an MPS with large sites.
Specifically, we compare timings for \ce{N2} 
with MPS-MRCISD based on a valence CAS(10e,8o), double and triple $\zeta$ bases and different maximal bond dimensions. 

The computations are performed with shared-memory parallelism\cite{zhai2021Low}
on one node with 28 Intel(R) Xeon(R)  E5-2680 v4 CPUs.
For each experiment, we start with a random initial state and perform two sweeps with the one-site DMRG algorithm and with perturbative noise.\cite{white2005noise,hubig2015noise}
For all computations, we measure and compare the total runtimes (including initialization steps such as the setup of the Hamiltonian matrix product operator).

\begin{figure*}[!htb]
  \includegraphics[width=.7\textwidth]{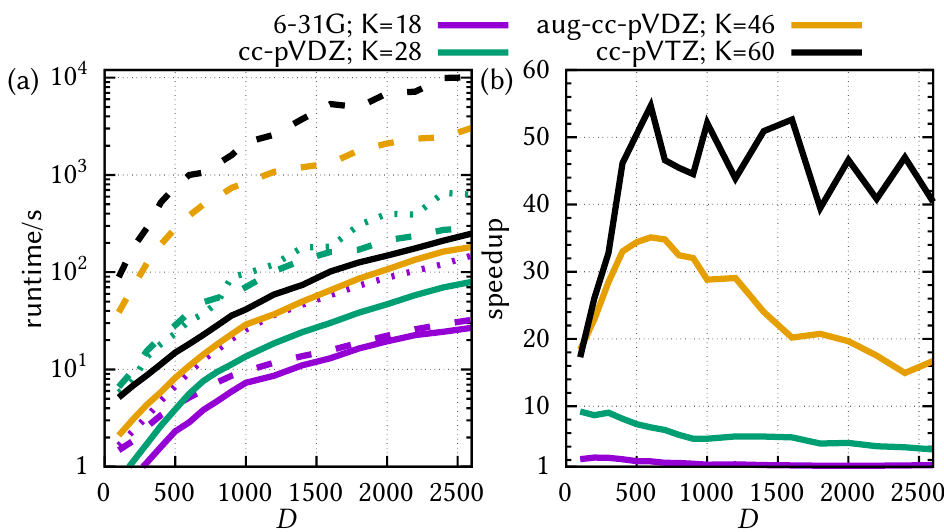}
    \caption{Runtime (left panel) and speedup (right panel) of different MPS-based simulations for \ce{N2} in different bases (total number of orbitals, $\Norb$ are shown).
    In the left panel, 
    the dashed %
    lines indicate  runtimes for an MPS-MRCISD computation with an MPS based on restricted quantum numbers (as many sites as orbitals).
    The full %
    lines indicate runtimes for the same MRCI setup but with an MPS where the virtual orbitals are collectively described by one large MPS site. 
    The MPS-MRCI computation is based on a valence CAS(10e,8o).
    The dotted lines indicate runtimes for a normal MPS/DMRG computation, approximating full CI.
    The right panel shows the speedup of using MPS-MRCI with a large-site MPS over using an MPS with restricted quantum numbers. 
    }
  \label{fig:N2_timing}
\end{figure*}

The absolute and relative timings are shown in \autoref{fig:N2_timing}.
For the same bond dimension, the MPS-MRCI simulations (full and dashed lines) are typically faster than the FCI-based DMRG simulations (dotted lines). 
This is expected as the Hilbert space size (and thus the matrices in the MPS) are restricted in the MPS-MRCI computation. 
Likewise, MPS-MRCI simulations converge faster than conventional MPS (FCI) simulations with respect to bond dimension (see below). %
For all basis sizes the large site MPS (full lines) performs significantly faster than the MPS based on restricted quantum numbers (dashed lines) (right panel in \autoref{fig:N2_timing}). 
The speedup is always significantly larger than 1, decreases with increasing bond dimension, and increases with basis size. 
For example, in a triple $\zeta$ basis with 60 orbitals, a speedup of more than 50 compared to the standard MPS can be obtained.
Notably, the large site MPS computation with a triple $\zeta$ basis (60 orbitals; full black line) is faster than the MPS with restricted quantum numbers with a double $\zeta$ basis (28 orbitals; dashed green line).

\begin{figure}[!htb]
  \includegraphics[width=\columnwidth]{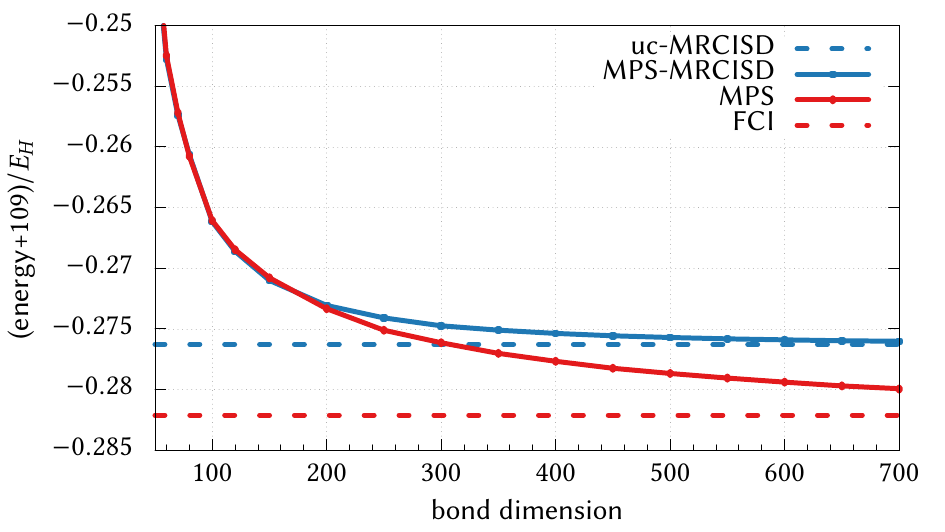}
  \caption{Bond dimension convergence for \ce{N2} at the equilibrium distance. The curves shown are MPS-MRCISD (blue curve) and normal DMRG (red curve). The dashed lines are the converged reference energies.} 
  \label{fig:n2_energy_bdim}
\end{figure}

The convergence of the energy versus bond dimension is shown 
for \ce{N2} in \autoref{fig:n2_energy_bdim}.
We compute the energy at the equilibrium distance ($R=\unit[1.1208]\angstrom$) using the cc-pVDZ basis. 
MRCISD is based on a full CAS(14e,10o), \new{employing a combination of natural orbitals obtained from CASSCF (for the CAS space) and from M\o{}ller-Plesset second order PT (MP2)\cite{helgaker_book} (by diagonalizing the one-particle density matrix in the space of the external orbitals). We use the same natural-occupation-based ordering for the MPS-MRCISD and the standard MPS wavefunctions.} %
While for this example the normal MPS approaches the uc-MRCISD energy more rapidly, namely around $D\sim300$, the overall convergence with respect to $\bdim$ is slower.
Due to the restrictions on the wavefunction, the MPS-MRCISD method requires a much smaller bond dimension of less than $400$ to approach an error of less than $\unit[1]{mE_H}$.
In contrast, a normal MPS requires a bond dimension of  $\sim\!\! 1000$ for a similar convergence tolerance.\cite{Stateoftheart2004chan}

\subsection{Chromium dimer}
\label{sec:appl_cr2}

The chromium dimer is a prototypical correlated system with complex bonding, 
which requires both a multireference treatment and a large amount of dynamic correlation.\cite{nevpt2001angeliLett,celani2004CIPT2,nevpt32006angeli,muller2009Cr2,kurashige2011caspt2,guo2016nev,vancoillie2016Cr2,luo2018ecMRCI,cr2AFQMC2015purwanto,li2020CrSHCI} 
Several studies have used both internally contracted and uncontracted MRCISD and related methods to compute the \ce{Cr2} binding curve.
Here, we will use the MPS-based formalism to obtain results for several variants of MPS-MRCI methods. Our purpose here is to illustrate the flexibility of the MPS formalism and the utility of the large site implementation which allows us to obtain results in large basis sets and beyond doubles excitations in the MRCI ansatz.

\begin{figure}[!htb]
  \includegraphics[width=\columnwidth]{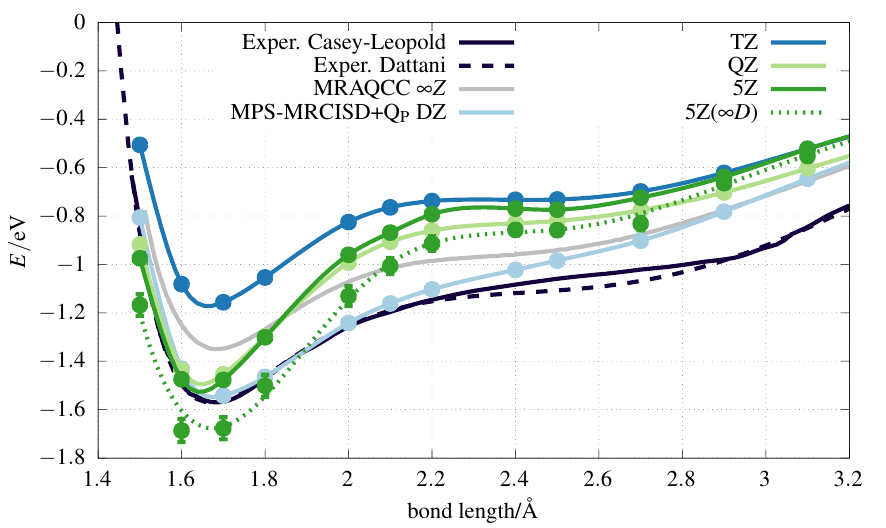}
    \caption{Potential energy curves of the chromium dimer. The black curves show experimental results from Casey and Leopold\cite{cr2PEC1993casey} (full lines) and a different fit to the same data by Dattani et al.\cite{cr22017Dattani} (dashed lines).
    The colored curves show 
    (uncontracted) MPS-MRCISD+Q$_\text{P}$ results for different basis sizes (see text for details). The 5$\zeta$ result is not fully converged with respect to bond dimension. An approximate extrapolation with bond dimension is shown as the green dotted curve.
    For reference, the gray curve shows  uncontracted MRAQCC(12,12) results in the basis set limit.\cite{muller2009Cr2} The MRAQCC PEC correlates fewer electrons and is based on a more restricted reference space. 
    }
  \label{fig:Cr2_mrci}
\end{figure}

We use a CAS self-consistent field  (CASSCF) reference with a valence CAS consisting of 12 electrons and 12 orbitals (3d and 4s shells, 28784 configuration state functions, CSFs) and employ the
spin-free exact-2-component Hamiltonian \cite{Exact2012peng,Quasirelativistic2005kutzelnigg}
with the cc-pV\new{$\{$D,T,Q,5$\}$}Z-DK basis sets (up to quintuple $\zeta$), which include up to $i$-type functions.\cite{dunningBas2006balabanov}
\new{To decrease the required bond dimension, we use CASSCF natural orbitals and Fiedler ordering in the active space.\cite{dmrgQI2011barcza,dmrgRev2015chan} 
We use standard canonical external and inactive orbitals since an MPS-MRCISD wavefunction is invariant with respect to orbital rotations in these spaces. }
We do not employ any BSSE correction.
The uncontracted MRCI wavefunction keeps the 1s, 2s, 2p shells frozen, and includes the 3s and 3p orbitals in the inactive space, and we correct the energies using the Pople, and other, size-extensivity  corrections. 
 (For comparison, previous uncontracted MRACPF and MRAQCC simulations by Dachsel et al.\cite{cr21999dachsel} and Müller\cite{muller2009Cr2}
used a generalized valence bond reference function consisting of 3088 or 1516 CSFs, %
using bases with up to $h$-type functions).
The large site representing the external space in the MPS has up to $153\cdot 10^3$ configurations. 
The PECs are generated from  
the binding energies as obtained by subtracting the energy from the dimer at large distance (to account for size consistency errors). Energy data is given in the Supporting Information.
The MPS-MRCI+Q$_\text{P}$ PECs are presented in \autoref{fig:Cr2_mrci}, together with the earlier uc-MRAQCC results from Müller,\cite{muller2009Cr2} and experimental curves (with a zero-point energy correction of $\unit[0.03]{eV}$~\cite{li2020CrSHCI}). 
The size consistency error at the 5$\zeta$ level is $\unit[0.185]{eV}$, which is similar to the uc-MRCI+Q results obtained by Müller.\cite{muller2009Cr2}
\new{As usual, we computed the size consistency error by taking the difference between the dimer energy at large distance and twice the energy of the $\ce{Cr}$ atom (based on restricted open-shell Hartree-Fock orbitals).}
We find that the MPS-MRCI simulations require a very large bond dimension,
typically in the middle of the MPS, which is larger than that required for the reference wavefunction.
\footnote{\new{The bond dimension at each site is restricted by a density matrix eigenvalue cutoff of $10^{-10}$ and by the restrictions of possible states that lie in each symmetry sector.}}
\new{The maximum bond dimension required in the MPS representation of the CAS(12,12) wavefunction is $\sim\!\!1,500$. In contrast, in the MPS-MRCI wavefunction, the  additional external space leads to a dramatic increase of the required bond dimension and}
with $\bdim=15,\!000$ (without spin adaptation) the Q$\zeta$ PEC is converged to the eye with accuracy of $\precsim\unit[1]{mE_H}$. 
\new{As mentioned in \autoref{sec:appl_N2}, the required bond dimension still is much smaller than that needed to represent the FCI wavefunction.}
However, we could not similarly converge the simulation with the 5$\zeta$ basis  using a maximum bond dimension of $16,\!000$. In particular, the relative energies for the bond lengths between $2.1$ and $\unit[2.5]\angstrom$ and the absolute energies are not converged at that bond dimension, thus  we also show an approximate extrapolation to infinite bond dimension\cite{dmrgRev2015chan}
for the 5$\zeta$ results.

Compared to the experimental curve, the MPS-MRCI+Q$_\text{P}$ PECs gives too narrow of a well and the $\sigma$-bonding around $\unit[2.2]{\angstrom}$ is underestimated.
The MPS-MRCI+Q results differ significantly from the earlier uc-MRAQCC PEC,\cite{muller2009Cr2} which mostly gives a qualitatively better curve, although the different size-extensivity corrections, basis sets, reference space, and number of correlated electrons makes it difficult to pinpoint the source of the difference. %

\begin{figure}[!htb]
  \includegraphics[width=\columnwidth]{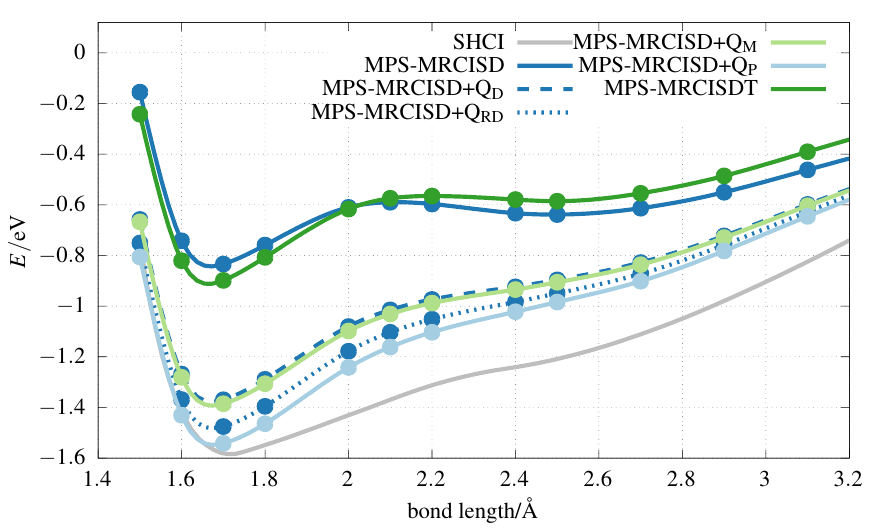}
    \caption{
    Potential energy curves of the chromium dimer in the cc-pVDZ-DK basis. Shown are MPS-MRCISD, MPS-MRCISD+Q,  and MPS-MRCISDT curves. The gray curve shows the selected heat bath configuration interaction (SHCI) result from Li et al.,\cite{li2020CrSHCI} which approximates the exact PEC in that basis.}
  \label{fig:cr2_mrci_nocorr}
\end{figure}

To see the effect of the size-extensivity correction on the curve shape, we also show the MPS-MRCISD+Q PECs with various corrections  for the cc-pVDZ-DK basis in \autoref{fig:cr2_mrci_nocorr}. Compared to the PEC without any correction (blue curve), the energy is shifted by $\sim\unit[0.8]{eV}$, illustrating the large error of uncorrected MRCISD.
All size-extensivity corrections lead to similar curve shapes, differing mostly in the energy shift. For the case of zero inactive orbitals, we found that the MPS-MRAQCC curve resembles the MRCI+Q$_\text{P}$ curve. 
Remarkably, the uncorrected MPS-MRCISD PEC leads to an additional minimum at larger bond distances, which is also the case for other methods such as valence-CAS-based CIPT2, CASPT2 and NEVPT3,\cite{muller2009Cr2,vancoillie2016Cr2,celani2004CIPT2} in particular if  small bases are used.
For MPS-MRCISD, we found the additional minimum to be more pronounced for larger bases. 
To estimate the effect of excitations beyond doubles, 
we additionally converged MPS-MRCISDT results with $\bdim=16,\!000$.
The MPS-MRCISDT PEC in the DZ basis \new{only has a very shallow additional minimum} and overall displays a qualitatively better PEC, albeit still very different from the more accurate 
selected heat-bath configuration interaction 
(SHCI) curve,\cite{li2020CrSHCI} which approximates FCI. 
These results indicate that the double minimum observed in MRCI treatments is mainly a size-consistency issue which is corrected by including higher order excitations explicitly, or a size-consistency correction, which partially accounts for the disconnected higher order pieces.

\subsection{Benzene}
\label{sec:appl_bz}

In a recent benchmark, the exact energy of the benzene molecule in the cc-pVDZ basis was approximated by a number of methods,\cite{eriksen2020bz} arriving at an estimated correlation energy of $\unit[-863.0]{mE_H}$ with an uncertainty of $\unit[\sim\!\!1]{mE_H}$. %
Here, we compare the accuracy of several uncontracted multireference methods implemented using the MPS within the large site formalism. We also provide the variational and extrapolated DMRG energy for comparison. (Results from other methods can be found in \lits{eriksen2020bz,Performance2020loos,Performance2020lee,Taming2021mahajan} but the DMRG estimate of error, which is an estimate of total error as a fraction of extrapolation error, is most directly comparable to the estimated errors reported here). 
All simulations use a valence CAS with 30 electrons in 30 orbitals, obtained from an MPS-CASSCF calculation.\cite{Density2008zgid,Orbital2008ghosh}
We use split-localized orbitals with $C_s$ symmetry, obtained from Edmiston-Ruedenberg localization,\cite{Localized1963edmiston}
followed by an additional DMRG-based internal orbital optimization.
The large site contains 78 \new{(canonical)} orbitals, resulting in $\sim\! 12\cdot 10^3$ configurations on the large site for the MPS-MRCISD-type of wavefunctions (as also used in MRPT2/3) and $\sim\! 630\cdot 10^3$ for an MPS-MRCISDT ansatz.

The energies of various methods, based on MPS-MRCISD, MPS-MRCISDT, MPS-MR perturbation theories and MPS-MRLCCM, are shown in \autoref{tab:bz}.
We make several observations. First, the total energies of the MPS-MRCI methods, with size-extensivity corrections, or even with triples, are quite poor.
This is likely due to the  large number of electrons, which \new{MRCI} methods were not designed to treat. 
Second, the MPS-MRREPT family of perturbation methods yields much better energies than the MPS-MRNEVPT family. 
MPS-MRREPT2 in particular yields surprisingly accurate energies, and is the only MPS-MR method to yield a more accurate estimate than CCSD(T) and CCSDT. 
 Finally, despite the similarity between the MPS-MRLCCM and MPS-MRREPT2 methods, %
they yield significantly different energies, illustrating the delicate balance needed when choosing the reference Hamiltonian in multi-reference perturbation theories.

\begin{table}[!htbp]
\caption{Correlation energies for benzene, computed using various (uncontracted) MPS multireference methods, based on a valence CAS(30, 30).
Shown are the maximal bond dimension ($\bdim$), the correlation energy ($\Delta E_\bdim$), the energy extrapolated to infinite bond dimension ($\Delta E$), and its error, defined as $1/5$ of its extrapolation distance.\cite{dmrgRev2015chan}
Note that the DMRG optimization from \lit{eriksen2020bz} was performed with spin symmetry included whereas all MPS-MR-based optimizations were performed without spin symmetry. Thus the MPS-MR methods require a larger bond dimension. 
The correlation energy is estimated to be around $\unit[-863]{mE_H}$.\cite{eriksen2020bz}}
\label{tab:bz}
\begin{tabular}{lrccc}
\toprule
  method & $\bdim$ &  \multicolumn{1}{c}{$\Delta E_\bdim/\unit{mE_H}$} &  \multicolumn{1}{c}{$\Delta E/\unit{mE_H}$}& \multicolumn{1}{c}{error /$\unit{mE_H}$}\\
  \midrule \\
CCSD(T)&& & -859.5 \\ %
CCSDT &&  & -859.9 \\ %
CCSDTQ&&  & -862.4 \\
DMRG   &6000& -859.2 & -862.8&0.7 \\
\midrule
MPS-CASSCF(30,30) & 4000 & -393.3\\
MPS-MRCISD & 9000 & -808.3 & -819.8 & 2.3\\
MPS-MRCISD+Q$_\text{RD}$ & 9000 & -864.8 & -880.7 & 3.0 \\
MPS-MRCISD+Q$_\text{P}$ & 9000 & -868.0 & -884.5 & 3.3\\
MPS-MRCISD+Q$_\text{M}$ & 9000 & -857.4 & -872.7 & 3.1\\
MPS-MRACPF & 9000 & -869.6 & -891.7 & 4.4\\
MPS-MRAQCC & 9000 & -864.0 & -875.5 & 2.3\\
MPS-MRCISDT & 9000 & -822.5 & -832.5 & 2.0 \\
MPS-MRREPT2 & 10000 & -857.6 & -862.0 & 0.9\\
MPS-MRREPT3 & 10000 & -850.1 & -854.5 & 0.9\\
MPS-MRNEVPT2 & 10000 & -779.3 & -783.0 & 0.7\\
MPS-MRNEVPT3 & 10000 & -829.9 & -834.3 & 0.9 \\
MPS-MRLCCM &  9000 & -872.9 & -889.9  & 3.4\\
\bottomrule
\end{tabular}
\end{table}
\section{Conclusions}
\label{sec:conclusions}

In summary, we have explored the advantages and disadvantages of clustering groups of orbitals into large sites in a matrix product state (MPS) from a computational perspective.
While often attractive from a chemical perspective, in many situations clustering leads to an increase in cost because of (1) the underlying exponential scaling of the cluster Hilbert space with cluster size and (2) longer-range inter-cluster correlations, which do not allow for a significant decrease in the MPS bond dimension.

A special case however is the MPS with large cluster sites at either end of the MPS. Because each large site only has  a single boundary (and does not have a boundary with another large site), when combined with a  configuration selection of the large site Hilbert space, there is a large regime of computational advantage. Here we explore the approximation of uncontracted multireference correlation theories using such a large site MPS.
We found that the large site MPS formalism yields  significant (more than an order of magnitude) computational speedups, compared to a conventional MPS implementation, for multireference wavefunctions with up to three particles in the external space. %
 General multireference theories are easily realized in this language, as we demonstrate by implementing multireference configuration interaction  (MRCI) and MRCI-based size-extensivity-corrected functionals such as averaged quadratic coupled cluster (MRAQCC),
various perturbation theories (PTs) such as $n$-electron valence-state PT (NEVPT) and restraining the excitation degree PT (REPT), and the multireference linearized coupled cluster method (MRLCCM).

We use the large site MPS implementation of the above theories to investigate some of the properties of the various multireference treatments in (a) the  nitrogen dimer, (b) the  chromium dimer, and (c) the  benzene molecule. 
Our computations used active spaces with up to
30 electrons and 30 orbitals, with up to triple excitations in the external space, and with up to 280 external orbitals. 
For \ce{Cr2}, our results show that the often observed double minimum in the potential energy curve is a result of the neglect of beyond double excitations, and can be corrected by their disconnected component (e.g. via size-consistency corrections to MRCI) or the explicit inclusion of triples. 
For the benzene molecule, we found that among the various theories mentioned, only the multireference REPT2 energy is within $\unit[1]{mE_H}$ of the estimate of the exact correlation energy. All other theories, including various size-extensivity-corrected MRCISD variants, MRCISDT, MRNEVPT, and MRLCCM, yield poor results showing (1) the need for size-extensive methods and (2) the importance of the choice of the reference Hamiltonian in multireference perturbation theories.

\if\USEACHEMSO1
\begin{acknowledgement}
\else
\acknowledgements
\fi
We thank Cyrus Umrigar for providing us with the SHCI PEC data for the chromium dimer. 
Work by GKC was supported by the US National Science Foundation (NSF) via grant no.~CHE-2102505. GKC acknowledges additional support from the Simons Foundation via the Many-Electron Collaboration and the Investigator Award. 
Work by HZ was supported by the US Department of Energy, Office of Science, Basic Energy Sciences under Award DE-SC0019374. 
Work by HRL was supported by the Air Force Office of Scientific Research, under Award FA9550-18-1-0095. 
HRL acknowledges support from a postdoctoral fellowship from the German Research Foundation (DFG) via grant LA 4442/1-1 during the first part of this work. 
\new{KG acknowledges support from a Ph.D. fellowship and a travel grant for a long stay abroad at the California Institute of Technology from the Research Foundation Flanders (FWO Vlaanderen).}
Some of the computations presented here were conducted at the Resnick High Performance Computing Center, a facility supported by the Resnick Sustainability Institute at the California Institute of Technology.
\if\USEACHEMSO1
\end{acknowledgement}
\fi

\appendix
\section{Cluster matrix product state for a hydrogen chain}
\label{sec:appl_H8}

Here, we show results for a \ce{H10} chain, 
in a cc-pVDZ basis\cite{ccpvdz} 
using a cluster MPS with selected configurations within each cluster. 
The linear nature of \ce{H10} chains make them very favorable for a description by an MPS, as well as for MPS-based clustering, since the average inter-cluster distance is large. 
We first discuss a chain of equidistantly spaced \ce{H} atoms with an atom separation of either $\unit[1.1]{a_0}$ (more delocalized) or $\unit[1.1]{\angstrom}$ ($\unit[2.08]{a_0}$; more insulating).
In the thermodynamic limit, the metal-insulator transition is  close to $\unit[1.7]{a_0}$, thus the more widely separated system is deep in the insulating regime. 
The shape of the orbitals is crucial for an efficient cluster MPS. %
To minimize entanglement, we use a localized basis obtained by 
aligning \ce{H10} along the $x$ axis and using orbitals that diagonalize $\hat x$. 
These localized orbitals were then grouped into 10 clusters, corresponding to the 10 \ce{H} atoms. 
To reduce the required number of configurations,
within each cluster, natural orbitals were
obtained by diagonalizing the
MP2 %
one-particle density matrix within the cluster block. 

The optimal selection of configurations in a large site can be defined from the exact wavefunction, computing the density matrix of the cluster in the configuration (e.g., determinantal) basis, and selecting those corresponding to the largest diagonal elements of the density matrix. 
Results obtained this way by first approximating the exact wavefunction for the full problem (using the variant of SHCI implemented in PySCF) are denoted "selection based on the full system".
We  also used a more scalable method where the density matrix of the cluster is approximated using a one-shot two-site density matrix embedding.\cite{dmet2012knizia,dmet2013knizia}
Here, a cluster and a neighboring cluster are chosen as the fragment and the remaining sites are represented by the density matrix embedding bath. The fragment plus bath problem is then solved via SHCI\cite{HeatBath2016holmes}.
Finally, the diagonal of the fragment density matrix is constructed and the configurations corresponding to the largest elements of the density matrix are used for subsequent cluster MPS calculations. 

\begin{figure}%
  \includegraphics[width=\columnwidth]{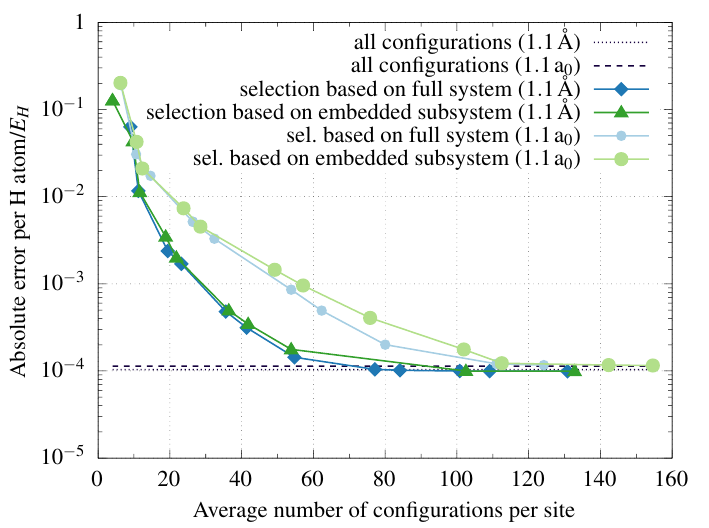}
  \caption{Convergence behavior of configuration selection in a cluster MPS for \ce{H10}/cc-pVDZ. 
  Shown is the absolute error per \ce{H} atom versus the average number of configurations per cluster/site. %
  The dark (pale) green and blue curves correspond to a 
  separation of the \ce{H} atoms of $\unit[1.1]{\angstrom}$ ($\unit[1.1]{a_0}$).
  The configuration selection is either based on the wavefunction from a selected heat-bath configuration interaction (SHCI) calculation of the full system (blue curves) or based on a SHCI calculation of each subsystem embedded in the full system (green curves); see text for details.
  The dotted (dashed) line corresponds to the error from a cluster MPS calculation with all 1024 configurations included for a separation of  $\unit[1.1]{\angstrom}$ ($\unit[1.1]{a_0}$). The remaining error is due to the finite bond dimension. 
  }
  \label{fig:h10_conf_sel}
\end{figure}

\autoref{fig:h10_conf_sel} compares the error in the energy as a function of included configurations per site. Compared to selection based on the full system,
the selection based on the embedded subsystems performs well.
To reach an accuracy of $\sim \unit[10^{-3}]{E_H}$
for a separation of $\unit[1.1]{a_0}$,
 $\Ndet\sim 50$ configurations per site are required, on average. This corresponds to $\sim 5\%$ of the total number of $1024$ possible configurations.
In the more stretched geometry with a separation of $\unit[1.1]{\angstrom}$, only around $30$ configurations are required for a similar accuracy.

\begin{figure}
  \includegraphics[width=\columnwidth]{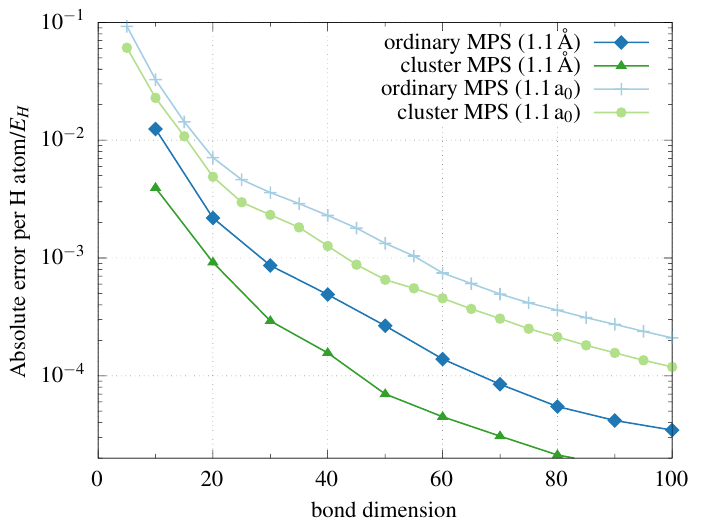}
  \caption{Convergence behavior of the bond dimension in  a cluster MPS (blue curves) versus standard MPS (green curves) for \ce{H10}/cc-pVDZ. The cluster MPS includes all possible configurations.
  The dark (pale) green and blue curves correspond to a 
  separation of the \ce{H} atoms of $\unit[1.1]{\angstrom}$ ($\unit[1.1]{a_0}$).
  }
  \label{fig:H10_bond_dim}
\end{figure}

To shed light on the possible reduction in bond dimension, we show a convergence plot in \autoref{fig:H10_bond_dim}.
Compared to an ordinary MPS, for a given error the bond dimension of the cluster MPS at $\unit[1.1]{a_0}$ is reduced by $\sim\!12$ -- $27\%$.
When the interatomic distance is changed from $\unit[1.1]{a_0}$ to $\unit[1.1]{\angstrom}$, the reduction in bond dimension increases to $\sim\!30$ -- $38\%$. 
Note that at the larger distance, far from the insulating transition, 
the atoms have a clear atomic character. The lack of a large reduction of bond dimension reflects the presence of longer-range correlations, in part from the long-range nature of the Coulomb interaction.

While the  small number of configurations per cluster is promising, the reduction in bond dimension is not. Comparing the scaling of an ordinary MPS with a cluster MPS, the computational effort is reduced by introducing clusters if 
$\Norb \bdim^3 > C \bdim_\text{cluster}^3 \Ndet$. For $\Ndet=50$, this would only be the case if the bond dimension of the cluster MPS, $\bdim_\text{cluster}$, were reduced by more than $54\%$.
For $\Ndet=30$, the reduction needs to be larger than $45\%$. Either requirement is not fulfilled at either of the two geometries.

\begin{figure}
  \includegraphics[width=\columnwidth]{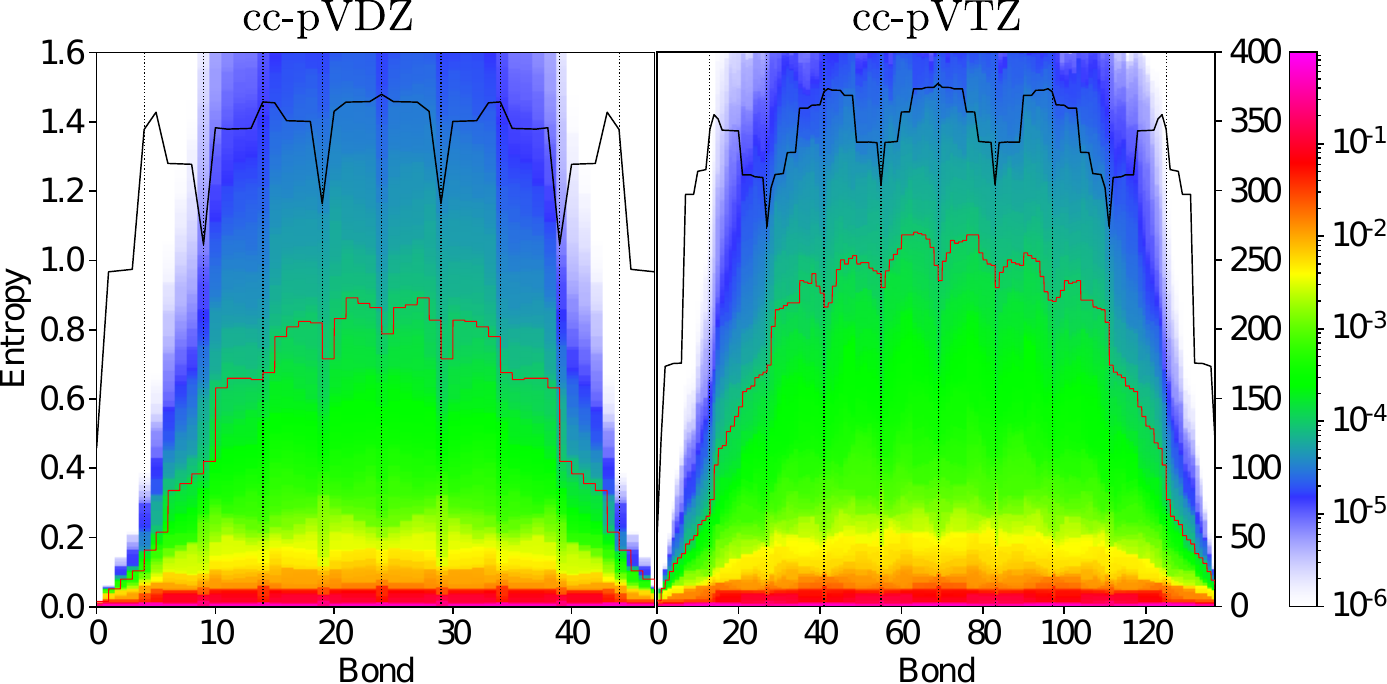}
  \caption{Singular values and von Neumann  entropy at different
  bipartitions in \ce{H10} (separation $\unit[1.1]{\angstrom}$) in a cc-pVDZ basis (left panel) and a cc-pVTZ basis (right panel).
  The colors corresponds to the magnitude of the singular values $\sigma_i$ at each bond.
  The red line denotes the contour at $\sigma = 10^{-4}$. The black line denotes the entropy at the corresponding bipartition.
  The dashed vertical lines denote the possible cluster decomposition.
  }
  \label{fig:H10_entropy}
\end{figure}

The required bond dimension in a cluster MPS can be better understood by analysing the singular values $\sigma_i$  of the site matrix and the corresponding  von Neumann entropy, $-\sum_i \sigma_i^2 \ln(\sigma_i^2)$, at each site in an MPS. 
The ideal case for a cluster MPS would be to have a very large entropy (large singular values) within each cluster but a small entropy (small singular values) at the boundaries of each cluster.
\autoref{fig:H10_entropy} shows this for \ce{H10} at a separation of $\unit[1.1]{\angstrom}$
in the cc-pVDZ and cc-pVTZ bases.\footnote{In contrast to the previous computations, here the localized basis has not been rotated into the natural orbital basis within each cluster. This, however, does not change the analysis of the bond dimension at the cluster boundaries.}
A decrease at the boundary of each cluster (vertical dashed lines) is only visible for some sites. 
In agreement with the results from \autoref{fig:H10_bond_dim}, the decrease is only marginal. 
Increasing the basis, i.e.~increasing the amount of long-range (dynamical) correlation that needs to be described, worsens the efficiency of clustering further.

\begin{figure}
  \includegraphics[width=\columnwidth]{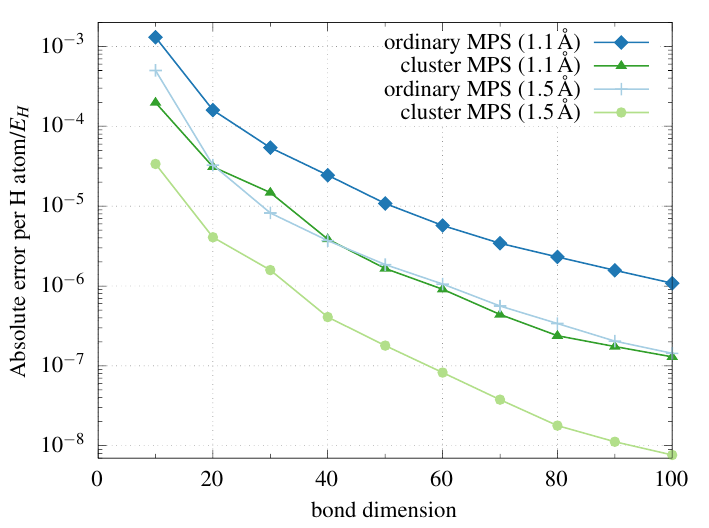}
  \caption{Same as \autoref{fig:H10_bond_dim} but showing 
  5 \ce{H2} molecules on a line, separated either $\unit[1.1]{\angstrom}$ (dark green/blue) or $\unit[1.5]{\angstrom}$ (pale green/blue) from each other.}
  \label{fig:H10_nonEq_bond_dim}
\end{figure}

\begin{figure}
  \includegraphics[width=\columnwidth]{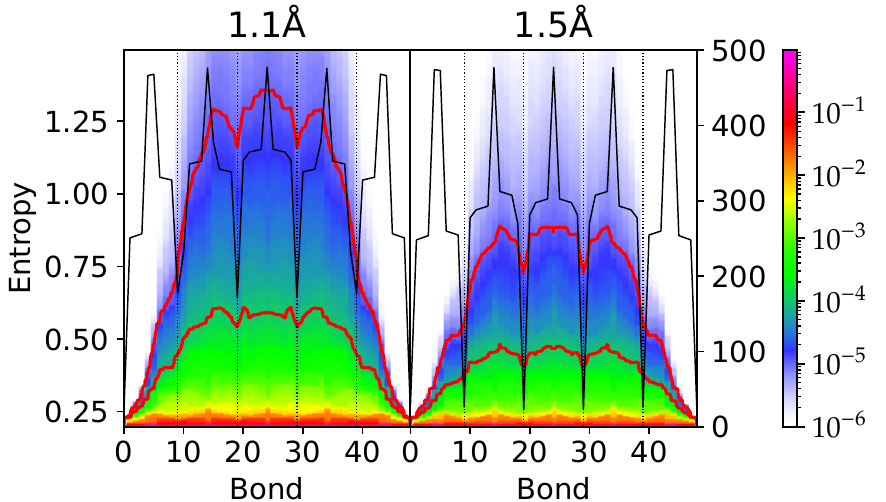}
  \caption{Same as \autoref{fig:H10_entropy} but  showing 
  5 \ce{H2} molecules on a line, separated by either $\unit[1.1]{\angstrom}$ (left panel) or $\unit[1.5]{\angstrom}$ (right panel) from each other. The red lines show a contour of the singular values at $10^{-5}$ (upper lines) and $10^{-4}$ (lower lines). We here show results for the cc-pVDZ basis.
 }
  \label{fig:H10_nonEq_entropy}
\end{figure}

We now discuss a \ce{H10} chain where 5 molecular \ce{H2} units (with bond distance of $\unit[0.74]\angstrom$, close to the molecular equilibrium geometry) are separated either by $\unit[1.1]\angstrom$ or  $\unit[1.5]\angstrom$ from each other.
Compared to the equidistant separation discussed above, these configurations have \ce{H2} as chemical units, thus there are 5 clusters instead of 10. 
The larger separation of the \ce{H2} units should favor clustering in the sense of a reduced bond dimension between the clusters, compared to the bond dimension within each cluster. 
\autoref{fig:H10_nonEq_bond_dim} compares the accuracy versus bond dimension for a normal and a cluster MPS. Note that, compared to equidistantly spaced \ce{H10},  the overall accuracy for a given bond dimension is much higher. At a separation of $\unit[1.5]\angstrom$  the bond dimension can be reduced by up to $\sim\!44$ -- $48\%$, indicating the more favorable clustering. At the smaller cluster distance of $\unit[1.1]\angstrom$ the decrease in bond dimension is reduced to $\sim\!36$ -- $44\%$. This is not much different from the equidistant H atom example discussed above. 
The singular values of the MPSs at each boundary are shown in \autoref{fig:H10_nonEq_entropy}. 
While the clustering is much more pronounced compared to \autoref{fig:H10_entropy}, the decrease of the bond dimension for a particular singular value at the boundary is not dramatic (red contour lines). Note that the first and the last \ce{H2} clusters in \ce{H10} cannot be resolved in the singular value plot.  
 
\new{We note that the simulations performed above are for one-dimensional problems, because this is the most favourable setting for the clustering approach. In 1D, the size of the cluster boundary does not scale with the size of the cluster, thus, for sufficiently weak interactions between the clusters, one can expect a regime where the inter-cluster interactions only generate a small number of excitations along the boundary independent of cluster size. The numerical simulations above, however, illustrate that this regime is not always reached in practice, in part due to the long-range of the Coulomb interaction. However, in say two-dimensional lattices (without additional structure) mapped onto one-dimensional slices, then even with local interactions, size-consistency dictates that one needs to retain both an exponential number of configurations per slice and an exponentially growing bond dimension as the system width increases, thus clustering is always asymptotically worse than the standard MPS approach. (See Supporting Information for additional numerical simulations on a 2D hydrogen lattice to illustrate this). Such general conclusions do not change either under a change of basis (e.g.~to a split-localized basis), which simply lead to different constants in the scaling, or (in the case of delocalized bases) worse asymptotic behaviour with system size.}


\new{
\section*{Supporting information}
See supporting information for  \ce{Cr2} energy data (Section I) and for data on the performance of a cluster MPS for a 4x4 hydrogen lattice.
}

\end{document}